\documentclass[]{iopart}
\usepackage[dvips]{epsfig}
\usepackage{subfigure}
\usepackage{float}
\usepackage{amsfonts}
\usepackage{amsmath}
\usepackage{amssymb}
\usepackage{rotating}
\usepackage{euscript}
\usepackage{subfigure}
\usepackage{enumerate}
\usepackage{hhline}
\usepackage{threeparttable}
\usepackage{supertabular}
\usepackage{multirow}
\usepackage{array}
\usepackage{tabularx}
\usepackage{rotating}

\newcommand{\vecb}[1]{\mathbf{#1}}
\newcommand{\rvecb}[1]{\mathbf{#1}}

\newcolumntype{Y}{>{\centering\arraybackslash}X}

\usepackage{iopams}  

\begin{document}
\bibliographystyle{osa}

\title[K. Srinivasan and O. Painter, Design of two-dimensional photonic crystal...] {Design of two-dimensional photonic crystal defect states for quantum cascade laser resonators}

\author{Kartik Srinivasan\footnote[3]{To whom correspondence should be addressed (kartik@caltech.edu)} and Oskar Painter\dag}

\address{\dag\ Thomas J. Watson, Sr. Laboratory of Applied Physics, California Institute of Technology, Pasadena, CA 91125, USA}

\date{\today}

\begin{abstract}
Current quantum cascade lasers based upon conduction band electron transitions are predominantly TM (electrical field normal to the epitaxial direction) polarized.  Here we present a study of localized defect modes, with the requisite TM polarization, in connected square and hexagonal lattice two-dimensional (2D) photonic crystals for application as quantum cascade laser resonators.  A simple group-theory based analysis is used to produce an approximate description of the resonant modes supported by defects situated at different high symmetry points within the host photonic lattice.  The results of this analysis are compared with 2D finite-difference time-domain (FDTD) simulations, showing a close correspondence between the two analyses, and potential applications of the analysis in quantum cascade laser design are considered.      
\end{abstract}



\section{Introduction}

Optical microcavities have been the subject of much research within a variety of fields, including optoelectronics and laser physics\cite{ref:Yokoyama2}, nonlinear optics\cite{ref:Spillane1}, cavity quantum electrodynamics\cite{ref:Kimble2}, and biomolecular sensing\cite{ref:Norris}.  Planar photonic crystal (PC) defect microcavities\cite{ref:Foresi,ref:Painter12}, formed by an intentionally introduced perturbation to the periodic photonic lattice, have attracted attention for a variety of reasons, including lithographic control of many salient properties of the cavity modes\cite{ref:Painter12,ref:Park1}, the potential for on-chip integration with PC waveguides\cite{ref:Noda2}, and their ability to support resonant modes that simultaneously exhibit ultra-small (wavelength-scale) modal volumes and high quality factors\cite{ref:Akahane2,ref:Srinivasan7}.  In most of these works, a two-dimensional (2D) connected dielectric lattice of air holes is employed, often as a result of experimental constraints (for example, a connected lattice is necessary for undercut membrane geometries).  The focus of these investigations has typically been on cavity modes with transverse electric (TE) polarization (magnetic field normal to the slab) in a hexagonal lattice, primarily because of the presence of a complete (in-plane) frequency band-gap for modes of such polarization\cite{ref:Joannopoulos}.  Recent calculations\cite{ref:Srinivasan1,ref:Ryu3} and experiments\cite{ref:Srinivasan7}, however, have confirmed that the absence of a complete in-plane band-gap (in these cases, for TE modes within a connected square lattice of air holes) does not preclude the formation of low-loss, localized resonant states, provided that the modes are localized to regions of momentum space where the lattice provides strong optical feedback.  These results thus provide motivation for the study of localized transverse magnetic (TM) polarized (electric field normal to the slab) states within connected photonic lattices, which also lack a complete in-plane band-gap.  

Further encouragement is found in recent work incorporating a PC microcavity within a quantum cascade (QC) semiconductor heterostructure\cite{ref:Faist} to form an electrically-injected PC QC laser\cite{ref:Colombelli2,ref:Srinivasan6}.  The intersubband transitions responsible for photon generation within QC lasers are fundamentally TM in nature, while efficient electrical injection necessitates the use of a connected PC lattice.  Though initial devices more employed more extended \emph{microcavity band-edge states}, future efforts will involve optimization of current injection to create further localized electrically-driven \emph{defect mode} lasers, attractive due to the potential for reduced cavity radiation loss (high quality factor) and further suppression of unwanted spontaneous emission.  For such applications, an understanding of the salient properties of the TM defect states will be valuable.  

Our study involves two complementary techniques for understanding the properties of the TM-polarized defect states in square and hexagonal lattice 2D PCs.  The first, detailed in Section \ref{sec:GT}, is a symmetry-based analysis analogous to that considered in previous articles\cite{ref:Painter12,ref:Painter13} for TE-like modes, where the tools of group theory are used to classify the symmetries and identify the dominant Fourier components of the defect states.  2D finite-difference time-domain (FDTD) simulations of the defect modes are presented in Section \ref{sec:FDTD_init}, and show a close correspondence with the group theory analysis, while complementing the results with a depiction of the real-space (in-plane) properties of the modes.  The results of these sections provide a starting point from which more advanced defect designs can be generated, and a simple example of this is given in Section \ref{sec:summary}.

\section{Symmetry Analysis}
\label{sec:GT}

The design of PC defect cavities has primarily been done through numerical methods such as FDTD.  The overriding role that symmetry plays in physical systems suggests that the translational and rotational symmetries present in photonic lattices should play a leading part in determining the properties of the defect modes, and as a result approximate, less numerically intensive methods can also provide valuable information.  In recent work\cite{ref:Painter12, ref:Painter13}, the methods of group theory were used to analyze and classify the modes of 2D PC slab WG defect cavities for TE-like modes.  The results of the symmetry-based analysis were compared with FDTD simulations and photoluminescence measurements, and showed a close correspondence between the approximate analysis, the detailed numerical simulations, and the experimental results.  In this section, we extend the previously developed analysis to the case of TM-like modes, with a focus on QC laser applications. 

The discrete symmetries of the dielectric function, $\epsilon(\vecb{r})$, determine the symmetries present within Maxwell's equations.  The group theory analysis we consider consists of two main steps.  Modes of the unperturbed slab waveguide are used as a symmetry basis to generate approximate field patterns for the PC modes at the high symmetry points within the first Brillouin zone (IBZ) of the reciprocal lattice.  The curvature of the photonic bands at these points are such that peaks and valleys are created in the energy-momentum dispersion surface.  It is these peaks and valleys which are the seeds from which localized states are formed.  The second step in our approach then utilizes these PC band-edge states as a basis to generate approximate forms for the localized defect modes lying within the (partial) band-gap.  

We consider quasi-2D PC structures in this paper, where an effective index ($n_{\text{eff}}$) for the fundamental TM mode is used to take into account vertical waveguiding.  In this quasi-2D system, the fundamental TM mode is represented by the scalar field $E_{z}$; in a true three dimensional (3D) PC slab WG system, modes can not be separated into pure TE and TM polarizations, but are nevertheless (for the lower lying frequency bands) significantly TE-like or TM-like in character.  As a result, the approximate analysis presented here will have utility in studies of 3D systems as well (indeed, in previous work\cite{ref:Painter12,ref:Painter13}, the close correspondence between the symmetry-based analysis and 3D FDTD results for TE-like modes was established).      

\subsection{Hexagonal Lattice}
\label{subsec:hex_lattice}

\begin{figure}[ht]
\begin{center}
\epsfig{figure=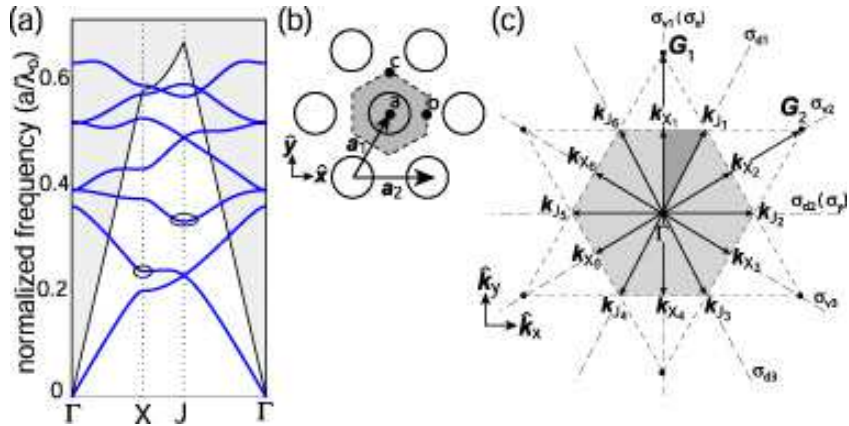, width=0.85\linewidth}
\caption{(a) Fundamental TM-like guided mode bandstructure ($n_{\text{eff}}=3.345$) for a hexagonal photonic lattice with $r/a$=0.35.  The air (cladding) light line is shown as a solid black line.  The X and J-point regions from which donor-like defect modes are formed are circled.  (b) Illustration of the (upper) real and (lower) reciprocal spaces of the two-dimensional hexagonal PC.  The high-symmetry points of the hexagonal lattice, referenced to the center of an air hole, are $\vecb{a}=(0,0)$, $\vecb{b}=(a/2,0)$, and $\vecb{c}=(0,a/\sqrt{3})$.}
\label{fig:hex_TM_band_and_lattice} 
\end{center}
\end{figure}       

The point group symmetry for TM modes in a 2D hexagonal lattice PC is $C_{6v}$.  A plot of the approximate in-plane bandstructure (calculated by the plane-wave expansion technique\cite{ref:Sakoda9}) for the fundamental TM-like guided modes in this lattice is given in Figure \ref{fig:hex_TM_band_and_lattice}(a), where $n_{\text{eff}}=3.345$ in the dielectric regions, and the ratio of the hole radius $r$ to lattice spacing $a$ of $r/a$=0.35 has been chosen.  

For the fundamental TM modes of an unpatterned waveguide, the electric field within the mirror plane of the slab is given by $\vecb{E}_{\rvecb{k_{\perp}}}(\vecb{r}_{\perp})=\hat{z}e^{-i(\rvecb{k_{\perp}}\cdot\vecb{r}_{\perp})}$, where $\rvecb{k}_{\perp}$ and $\vecb{r}_{\perp}$ are in-plane wavenumber and spatial coordinates, respectively.  Upon patterning the slab waveguide, coupling occurs between waveguide modes with similar unperturbed frequencies and propagation constants modulo $\rvecb{G}$, where $\rvecb{G}$ is a reciprocal lattice vector.  This follows from the (approximate) conservation of frequency and crystal momentum.

The high symmetry points of the first Brillouin zone (IBZ) of the hexagonal lattice (Figure \ref{fig:hex_TM_band_and_lattice}) are the six degenerate $X$-points ($\{\pm(0,1)k_X$, $\pm(\sqrt{3}/2,1/2)k_X$, $\pm(\sqrt{3}/2,-1/2)k_X\}$), the six degenerate $J$-points ($\{\pm(1/2,\sqrt{3}/2)k_J$, $\pm(1/2,-\sqrt{3}/2)k_J$, $\pm(1,0)k_J\}$), and the $\Gamma$-point=$(0,0)$, with $k_X=2\pi/a\sqrt{3}$ and $k_J=4\pi/3a$. The group of the wave vector, which defines the point group symmetry of a plane wave modulo $\rvecb{G}$ within the dielectric lattice, is $C_{2v}$, $C_{3v}$, and $C_{6v}$ for the $X$, $J$, and $\Gamma$ points, respectively.  Character tables\cite{ref:Tinkham} for these groups are given in Table \ref{table:PG_char_table}. 

Unlike the case for TE modes, the TM mode bandstructure does not exhibit a clear in-plane bandgap.  Nevertheless, the curvature of the photonic bands is such that we can still expect localized defect states to form, provided that the modes are composed of Fourier components that lie in a region of momentum space where a bandgap does exist (i.e., a partial in-plane bandgap).  In particular, the $J$-point of the third band (by frequency) has the correct curvature to form donor-type modes, and enough of a bandgap surrounding it to create localized states (the $X$-point of the second band also has the correct curvature, but because the bandgap surrounding it is somewhat small we will not consider it).

\begin{table}
\caption{Point Group character tables for the hexagonal lattice.}
\begin{center}
\begin{tabular}{ll}
\multicolumn{2}{l}{\begin{tabular}{c|rrrrrr||c|rrrr}
\hline
$C_{6v}$&$E$&$C_2$&$2C_3$&$2C_6$&$3\sigma_d$&$3\sigma_v$ & $C_{2v}$&$E$&$C_2$&$\sigma_x(\sigma_{v1})$&$\sigma_y(\sigma_{d2})$ \\
\hline
$A_1^{\prime\prime}$ & $1$ & $1$ & $1$ & $1$ & $1$ & $1$ & $A_1$ & $1$ & $1$ & $1$ & $1$ \\
$A_2^{\prime\prime}$ & $1$ & $1$& $1$ & $1$ & $-1$ & $-1$ & $A_2$ & $1$ & $1$ & $-1$ & $-1$  \\
$B_1^{\prime\prime}$ & $1$ & $-1$ & $1$ & $-1$ & $1$ & $-1$ & $B_1$ & $1$ & $-1$ & $-1$ & $1$ \\
$B_2^{\prime\prime}$ & $1$ & $-1$ & $1$ & $-1$ & $-1$ & $1$ & $B_2$ & $1$ & $-1$ & $1$ & $-1$ \\
$E_1$ & $2$ & $-2$ & $-1$ & $1$ & $0$ & $0$ & \\
$E_2$ & $2$ & $2$ & $-1$ & $-1$ & $0$ & $0$ & \\
\hline
$S^{a,d_{J3}}$ & $2$ & $0$ & $2$ & $0$ & $2$ & $0$ & $S^{a,d_{J3}}$ & $2$ & $0$ & $0$ & $2$ \\
           &     &     &     &     &     &     & $S^{b,d_{J3}}$ & $2$ & $0$ & $0$ & $2$ \\
\end{tabular}} \\

\\

\multicolumn{2}{l}{\begin{tabular}{c|rrr||c|rrr}
\hline
$C_{3v,\sigma_d}$&$E$&$2C_3$&$3\sigma_d$ & $C_{3v,\sigma_v}$&$E$&$2C_3$&$3\sigma_v$ \\
\hline
$A_1^\prime$ & $1$ & $1$ & $1$ & $A_1^{\prime\prime\prime}$ & $1$ & $1$ & $1$ \\
$A_2^\prime$ & $1$ & $1$ & $-1$ & $A_2^{\prime\prime\prime}$ & $1$ & $1$ & $-1$ \\
$E$ & $2$ & $-1$ & $0$ & $E$ & $2$ & $-1$ & $0$ \\
\hline
    &     &      &     & $S^{c,d_{J3}}$ & $3$ & $0$ & $-1$ \\
\end{tabular}} \\

\end{tabular}
\end{center}
\label{table:PG_char_table}
\end{table}

\subsubsection{J-point Band-Edge Representation}
\label{subsubsec:hex_J_point}
In determining an approximate description of the modes of the patterned slab waveguide we consider those modes of the unpatterned slab waveguide that are most strongly coupled considering temporal and spatial frequency conservation.  At the $J$-point these modes have a scalar electric field (TM) given by $\vecb{E}=\hat{z}e^{-i\rvecb{k}_{J_i}\cdot \vecb{r_{\perp}}}$ and a frequency $\omega_o^J \approx c|\rvecb{k}_{J}|/n_{\text{eff}}$.   

The star of $\rvecb{k}$ ($\star\rvecb{k}$) at the $J$-point, which consists of all the independent $J$-points within the IBZ, is given by (not uniquely) $\{\rvecb{k}_{J_1},\rvecb{k}_{J_2}\}$.  A symmetry basis for the modes of the patterned slab waveguide at the $J_1$-satellite point can be found by applying the symmetry operations of the group of the wave vector ($\mathcal{G}_{o\rvecb{k}_{J}}=C_{3v}$) to the seed vector $\vecb{E}_{\rvecb{k}_{J_1}}$.  In this case, the basis is $(\vecb{E}_{\rvecb{k}_{J_1}},\vecb{E}_{\rvecb{k}_{J_3}},\vecb{E}_{\rvecb{k}_{J_5}})$.  Projecting this symmetry basis onto the IRREP spaces of $C_{3v}$ yields:

\begin{equation}
\label{eq:hex_J_1_3_5_modes}
\begin{split}
\vecb{E}_{A_1^{\prime}} & =\hat{z}(e^{-i\rvecb{k}_{J_1}\cdot\vecb{r}^{a}_{\perp}}+e^{-i\rvecb{k}_{J_3}\cdot\vecb{r}^{a}_{\perp}}+e^{-i\rvecb{k}_{J_5}\cdot\vecb{r}^{a}_{\perp}}), \\
\vecb{E}_{E,1} & =\hat{z}(e^{-i\rvecb{k}_{J_1}\cdot\vecb{r}^{a}_{\perp}}+e^{-i\rvecb{k}_{J_3}\cdot\vecb{r}^{a}_{\perp}}-2e^{-i\rvecb{k}_{J_5}\cdot\vecb{r}^{a}_{\perp}}), \\
\vecb{E}_{E,2} & =\hat{z}(e^{-i\rvecb{k}_{J_1}\cdot\vecb{r}^{a}_{\perp}}-e^{-i\rvecb{k}_{J_3}\cdot\vecb{r}^{a}_{\perp}}),
\end{split}
\end{equation}

\noindent where $A_1^{\prime}$ and $E$ label the IRREP spaces of $C_{3v}$ (see Table \ref{table:PG_char_table}; $E$ is a 2D IRREP), and the index $a$ is used to denote the location of the origin at point $a$ in the hexagonal lattice.  Because the dielectric structure has a low index air hole at point $a$, we associate $\vecb{E}_{A_1^{\prime}}$ with the third highest frequency band mode.  This is a result of the tendency, all things being equal, for modes with electric field concentrated within regions of high dielectric constant to be lower frequency than those with electric field concentrated in low dielectric regions\cite{ref:Joannopoulos}.  We associate the set of modes \{$\vecb{E}_{E,1},\vecb{E}_{E,2}$\}, with the first and second frequency bands, which, consistent with the bandstructure calculations (Figure \ref{fig:hex_TM_band_and_lattice}(a)), are degenerate at the $J$-point.   

To fully define the modes at the $J$-point all modes of the $\star \rvecb{k}$ must be included.  The result of doing so is the following set of lower band modes,

\begin{equation}
\label{eq:hex_J_VB_modes_a}
J(1,2)_{a} = \hat{z} 
\begin{pmatrix}
e^{-i\rvecb{k}_{J_1}\cdot\vecb{r}^{a}_{\perp}}+e^{-i\rvecb{k}_{J_3}\cdot\vecb{r}^{a}_{\perp}}-2e^{-i\rvecb{k}_{J_5}\cdot\vecb{r}^{a}_{\perp}} \\
e^{-i\rvecb{k}_{J_1}\cdot\vecb{r}^{a}_{\perp}}-e^{-i\rvecb{k}_{J_3}\cdot\vecb{r}^{a}_{\perp}} \\
e^{-i\rvecb{k}_{J_2}\cdot\vecb{r}^{a}_{\perp}}+e^{-i\rvecb{k}_{J_4}\cdot\vecb{r}^{a}_{\perp}}-2e^{-i\rvecb{k}_{J_6}\cdot\vecb{r}^{a}_{\perp}} \\
e^{-i\rvecb{k}_{J_2}\cdot\vecb{r}^{a}_{\perp}}-e^{-i\rvecb{k}_{J_4}\cdot\vecb{r}^{a}_{\perp}}
\end{pmatrix},
\end{equation}

\noindent and higher band modes, 

\begin{equation}
\label{eq:hex_J_VB_modes_a_2}
J3_{a} = \hat{z} 
\begin{pmatrix}
e^{-i\rvecb{k}_{J_1}\cdot\vecb{r}^{a}_{\perp}}+e^{-i\rvecb{k}_{J_3}\cdot\vecb{r}^{a}_{\perp}}+e^{-i\rvecb{k}_{J_5}\cdot\vecb{r}^{a}_{\perp}} \\
e^{-i\rvecb{k}_{J_2}\cdot\vecb{r}^{a}_{\perp}}+e^{-i\rvecb{k}_{J_4}\cdot\vecb{r}^{a}_{\perp}}+e^{-i\rvecb{k}_{J_6}\cdot\vecb{r}^{a}_{\perp}} 
\end{pmatrix},
\end{equation}

The approximate $J$-point band-edge modes derived above have their origin at the center of an air hole (point $a$).  The hexagonal lattice has two other high-symmetry points around which one may center a defect, points $b$ and $c$ shown in Figure \ref{fig:hex_TM_band_and_lattice}(b).  Unlike point $a$, points $b$ and $c$ are of lower symmetry than that of the lattice ($C_{2v}$ and $C_{3v,\sigma_v}$, respectively).       

To determine the approximate band-edge modes for defect regions centered about these points, the lower and higher band modes are simply shifted to each of these points using the transformations $\vecb{r}^{b}_{\perp}=\vecb{r}_{\perp}-\vecb{b}$ and $\vecb{r}^{c}_{\perp}=\vecb{r}_{\perp}-\vecb{c}$.  The resulting bases are labeled $J(1,2)_{b}$, $J3_{b}$, $J(1,2)_{c}$,and $J3_{c}$ for lower and higher bands centered about points $b$ and $c$, respectively.     

\subsubsection{$J$-point Donor Modes}
\label{subsubsec:hex_c_donor}
We now consider donor modes, which are formed by a local increase in the dielectric constant within the photonic lattice.  As discussed previously, the lowest frequency region of $\rvecb{k}$-space for which this can occur, while still enjoying the benefits of significant isolation from other bands, is the $J$-point of the third frequency band.  Therefore, the appropriate symmetry basis to use for describing localized donor modes are the  band-edge modes of $J3_{a}$, $J3_{b}$, and $J3_{c}$ for defects centered around points $a$, $b$, and $c$, respectively.  For defects centered about point $a$ the largest possible symmetry is that of the underlying hexagonal lattice, $C_{6v}$, whereas for defects about point $b$ and point $c$ the largest point group symmetries are $C_{2v}$ and $C_{3v,\sigma_v}$, respectively.  Correspondingly, the character values of representation $S^{a,d_{J3}}$ of the $J3_{a}$ basis under $C_{6v}$, representation $S^{b,d_{J3}}$ of the $J3_{b}$ basis under $C_{2v}$, and representation $S^{c,d_{J3}}$ of the $J3_{c}$ basis under $C_{3v,\sigma_v}$ are given in Table \ref{table:PG_char_table}.  From these character tables we find that these representations decompose as $S^{a,d_{J3}}=A_1^{\prime\prime}\oplus B_1^{\prime\prime}$, $S^{b,d_{J3}}=A_1\oplus B_1$, and $S^{c,d_{J3}}=E \oplus A_2^{\prime\prime\prime}$.  Using the appropriate projection operators\cite{ref:Tinkham} on $J3_{a}$, a set of basis functions for the localized conduction band donor modes about point $a$ is found.  Similarly, projecting the basis functions $J3_{b}$ onto the IRREPs of $C_{2v}$ and the basis functions $J3_{c}$ onto the IRREPs of $C_{3v,\sigma_v}$, we get the donor modes about point $b$.  The approximate forms of the donor modes found from these projections are given in Table \ref{table:hex_lattice_TM_donor_modes}.  Also shown in this table is the new mode structure for a defect cavity of reduced symmetry ($C_{2v}$) centered about point $a$ (points $b$ and $c$ are already of lower point symmetry than the lattice), found by using the compatibility relations between the IRREPs of the full and reduced symmetry groups.

\renewcommand{\arraystretch}{1.2}
\renewcommand{\extrarowheight}{0pt}
\begin{table}
\caption{Symmetry and dominant Fourier components of the $\vecb{E}$-field for TM donor modes formed from the $J_{3}$ band-edge in a hexagonal lattice.}
\label{table:hex_lattice_TM_donor_modes}
\begin{tabular*}{\linewidth}{cc}
\hline
\hline
Cavity Center/Symmetry & Mode \\
\hline 
$a/C_{6v}$ & $\vecb{E}_{A_1^{\prime\prime}}^{a,d_{J3}} = \hat{z} \bigl(\cos(\rvecb{k}_{J_1}\cdot\vecb{r}^{a}_{\perp})+\cos(\rvecb{k}_{J_3}\cdot\vecb{r}^{a}_{\perp})+\cos(\rvecb{k}_{J_5}\cdot\vecb{r}^{a}_{\perp})\bigr)$ \\
$a/C_{6v}$ & $\vecb{E}_{B_1^{\prime\prime}}^{a,d_{J3}} = \hat{z} \bigl(\sin(\rvecb{k}_{J_1}\cdot\vecb{r}^{a}_{\perp})+\sin(\rvecb{k}_{J_3}\cdot\vecb{r}^{a}_{\perp})+\sin(\rvecb{k}_{J_5}\cdot\vecb{r}^{a}_{\perp})\bigr)$ \\
$b/C_{2v}$ & $\vecb{E}_{A_1}^{b,d_{J3}} = \hat{z} \bigl(\cos(\rvecb{k}_{J_1}\cdot\vecb{r}^{b}_{\perp})+\cos(\rvecb{k}_{J_3}\cdot\vecb{r}^{b}_{\perp})-\cos(\rvecb{k}_{J_5}\cdot\vecb{r}^{b}_{\perp})\bigr)$ \\
$b/C_{2v}$ & $\vecb{E}_{B_1}^{b,d_{J3}} = \hat{z} \bigl(\sin(\rvecb{k}_{J_1}\cdot\vecb{r}^{b}_{\perp})+\sin(\rvecb{k}_{J_3}\cdot\vecb{r}^{b}_{\perp})-\sin(\rvecb{k}_{J_5}\cdot\vecb{r}^{b}_{\perp})\bigr)$ \\
$c/C_{3v,\sigma_v}$ & $\vecb{E}_{E,1}^{c,d_{J3}} = \hat{z} \bigl(\cos(\rvecb{k}_{J_1}\cdot\vecb{r}^{c}_{\perp}+\frac{2\pi}{3})+\cos(\rvecb{k}_{J_3}\cdot\vecb{r}^{c}_{\perp}-\frac{2\pi}{3})$ \\
 & $+ \cos(\rvecb{k}_{J_5}\cdot\vecb{r}^{c}_{\perp})\bigr)$ \\
$c/C_{3v,\sigma_v}$ & $\vecb{E}_{E,2}^{c,d_{J3}} = \hat{z} \bigl(\sin(\rvecb{k}_{J_1}\cdot\vecb{r}^{c}_{\perp}+\frac{2\pi}{3})+\sin(\rvecb{k}_{J_3}\cdot\vecb{r}^{c}_{\perp}-\frac{2\pi}{3})$ \\
 & $+ \sin(\rvecb{k}_{J_5}\cdot\vecb{r}^{c}_{\perp})\bigr)$ \\
\hline
$a/C_{2v}$ & $\vecb{E}_{A^{\prime\prime}_1}^{a,d_{J3}}  \rightarrow  \vecb{E}_{A_1}^{a,d_{J3}}$ \\
$a/C_{2v}$ & $\vecb{E}_{B^{\prime\prime}_1}^{a,d_{J3}}  \rightarrow  \vecb{E}_{B_1}^{a,d_{J3}}$ \\  
\hline
\hline
\end{tabular*}
\end{table}
\renewcommand{\arraystretch}{1.0}

The key pieces of information given by our approximate description are the dominant Fourier components that comprise each mode.  Approximate forms for $\emph{localized}$ defect states can be produced by including some form of an envelope function along with these approximate symmetry-based forms; in previous work\cite{ref:Painter14}, the effects of the underlying photonic lattice are captured through derivation of a Wannier-like wave equation for the envelope of resonant TE modes.  When combined with the results of the group theory based analysis, a very accurate picture (as confirmed by FDTD simulations) of the near-field behavior of the defect modes is produced.  A similar Wannier-like equation for TM modes can be generated; however, for the purposes of this paper, we forego such an analysis and instead obtain more detailed information directly from FDTD simulations.  The key point stressed in Ref. \cite{ref:Painter14} that is of relevance here is that the ground state envelope functions transform effectively as the identity operator, and as a result have no bearing on the transformation properties (i.e., the classification of the symmetries of the modes) derived here. 

\subsection{Square Lattice}
\label{subsec:sq_lattice}
\begin{figure}[ht]
\begin{center}
\epsfig{figure=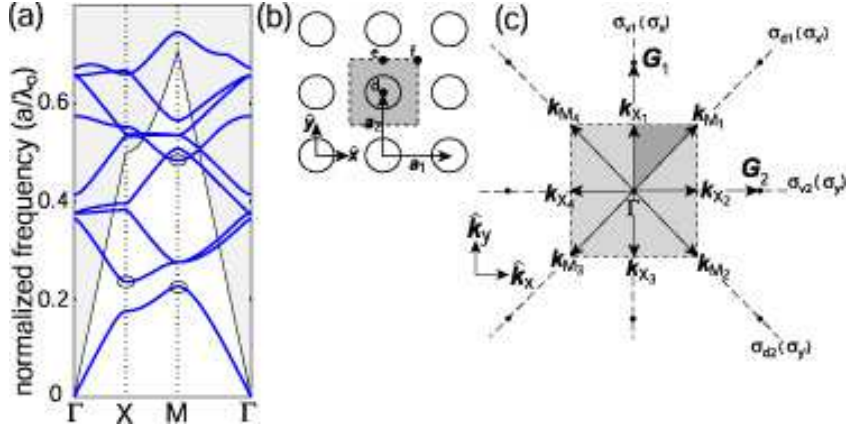, width=0.85\linewidth}
\caption{(a) Fundamental TM-like guided mode bandstructure ($n_{\text{eff}}=3.345$) for a square photonic lattice with $r/a$=0.40.  The air (cladding) light line is shown as a solid black line.  The X and M-point regions from which defect modes are formed are circled in black. (b) Illustration of the (upper) real and (lower) reciprocal spaces of the two-dimensional PC with a square array of air holes.  The high-symmetry points of the square lattice, referenced to the center of an air hole, are $\vecb{d}=(0,0)$, $\vecb{e}=(0,a/2)$, and $\vecb{f}=(a/2,a/2)$.}
\label{fig:sq_TM_band_lattice} 
\end{center}
\end{figure}

The point group symmetry of the square lattice photonic crystal is $C_{4v}$.  A plot of the approximate in-plane bandstructure for the fundamental TM-like guided modes of a square lattice of air holes with $n_{\text{eff}}=3.345$ for the dielectric regions and $r/a$=0.40 is given in Figure \ref{fig:sq_TM_band_lattice}(a).
    
The high symmetry points of the IBZ of the square lattice (Figure \ref{fig:sq_TM_band_lattice}(b)) are the four degenerate $X$-points ($\{\pm(1,0)k_X$, $\pm(0,1)k_X\}$), the four degenerate $M$-points ($(\pm\sqrt{2}/2$,$\pm\sqrt{2}/2)k_M$), and the $\Gamma$-point=$(0,0)$, with $k_X=\pi/a$ and $k_M=\pi\sqrt{2}/a$.  The group of the wave vector at the $X$-, $M$-, and $\Gamma$-points is $C_{2v}$, $C_{4v}$, and $C_{4v}$, respectively.  Character tables\cite{ref:Tinkham} for these groups are given in Table \ref{table:PG_char_table_2}. 

Based on the photonic bandstructure diagrams of Figure \ref{fig:sq_TM_band_lattice}(a), the candidate regions of $\rvecb{k}$-space from which we expect to form defect modes include the $X$-point off the second frequency band (conduction band donor modes) and the $M$-point off the fourth frequency band (higher band donor modes).  Discussion of the $M$-point of the first frequency band, also a candidate region of $\rvecb{k}$-space (to create valence band acceptor modes) is not considered here in the interests of keeping the discussion as concise as possible.  The analysis is quite similar to what was done for the hexagonal lattice; we begin by forming basis functions to describe the band-edge modes, and then couple these modes via the symmetry of the defect to get an approximate description of the cavity modes.  

\begin{table}
\caption{Point Group character tables for the square lattice.}
\begin{center}
\begin{tabular}{l}

\begin{tabular}{c|rrrrr||c|rrrr}
\hline
$C_{4v}$&$E$&$C_2$&$2C_4$&$2\sigma_v$&$2\sigma_d$&$C_{2v,\sigma_v}$&$E$&$C_2$&$\sigma_x(\sigma_{v1})$&$\sigma_y(\sigma_{v2})$\\
\hline
$A_1^{\prime\prime}$ & $1$ & $1$ & $1$ & $1$ & $1$ & $A_1$ & $1$ & $1$ & $1$ & $1$ \\
$A_2^{\prime\prime}$ & $1$ & $1$& $1$ & $-1$ & $-1$ & $A_2$ & $1$ & $1$ & $-1$ & $-1$ \\
$B_1^{\prime\prime}$ & $1$ & $1$ & $-1$ & $1$ & $-1$ & $B_1$ & $1$ & $-1$ & $-1$ & $1$ \\
$B_2^{\prime\prime}$ & $1$ & $1$ & $-1$ & $-1$ & $1$ & $B_2$ & $1$ & $-1$ & $1$ & $-1$ \\
$E$ & $2$ & $-2$ & $0$ & $0$ & $0$ & \\
\hline
$S^{M}$ & $4$ & $0$ & $0$ & $0$ & $2$ & $S^{X_1}$ & $2$ & $0$ & $2$ & $0$ \\
$S^{d,d_{X2}}$ & $2$ & $2$ & $0$ & $2$ & $0$ & $S^{e,d_{X2}}$ & $2$ & $0$ & $2$ & $0$ \\
$S^{d,d_{M4}}$ & $1$ & $1$ & $1$ & $1$ & $1$ & $S^{e,d_{M4}}$ & $1$ & $-1$ & $1$ & $-1$  \\
$S^{f,d_{X2}}$ & $2$ & $-2$ & $0$ & $0$ & $0$ & \\
$S^{f,d_{M4}}$ & $1$ & $1$ & $-1$ & $-1$ & $1$ & \\
\end{tabular}\\

\end{tabular}
\end{center}
\label{table:PG_char_table_2}
\end{table}

\subsubsection{X-point band-edge representation}
\label{subsubsec:sq_X_point}
A symmetry basis for the modes of the square lattice PC at the $X$-point is found by applying the symmetry operations of the group of the wave vector ($\mathcal{G}_{o\rvecb{k}_{X}}=C_{2v}$) to the seed vector $\vecb{E}_{\rvecb{k}_{X_1}}$.  In this case, the basis is simply $(\vecb{E}_{\rvecb{k}_{X_1}},\vecb{E}_{-\rvecb{k}_{X_1}})$.  Projecting this symmetry basis onto the IRREP spaces of $C_{2v}$, and then including all the modes of the star of $\star \rvecb{k}$, results in the following set of degenerate valence band modes,

\begin{equation}
\label{eq:sq_X_VB_modes_d}
X1_{d} = \hat{z}
\begin{pmatrix} 
\sin(\rvecb{k}_{X_1}\cdot\vecb{r}^{d}_{\perp}) \\
\sin(\rvecb{k}_{X_2}\cdot\vecb{r}^{d}_{\perp}) 
\end{pmatrix},
\end{equation}

\noindent and degenerate conduction band modes, 

\begin{equation}
\label{eq:sq_X_CB_modes_d}
X2_{d} =  \hat{z}
\begin{pmatrix} 
\cos(\rvecb{k}_{X_1}\cdot\vecb{r}^{d}_{\perp}) \\
\cos(\rvecb{k}_{X_2}\cdot\vecb{r}^{d}_{\perp}) 
\end{pmatrix},
\end{equation}

In the square lattice there are three different high-symmetry points around which one may center a defect, labeled $d$, $e$, and $f$ in Figure \ref{fig:sq_TM_band_lattice}(b).  Points $d$ and $f$ maintain the $C_{4v}$ symmetry of the lattice, while point $e$ has the lowered symmetry $C_{2v,\sigma_v}$.  The shifted valence and conduction band-edge modes are found through the transformations $\vecb{r}^{e}_{\perp}=\vecb{r}_{\perp}-\vecb{e}$ and $\vecb{r}^{f}_{\perp}=\vecb{r}_{\perp}-\vecb{f}$, and are labeled as $X1_{e}$, $X2_{e}$, $X1_{f}$, and $X2_{f}$ for points $e$ and $f$ respectively.  

\subsubsection{M-point band-edge representation}
\label{subsubsec:sq_M_point}
The group of the wavevector at the $M$-point is $C_{4v}$, and as such the symmetry basis, $S^{M}$, includes all the $M$-points of the IBZ, $S^{M}=(\vecb{E}_{\rvecb{k}_{M_1}},\vecb{E}_{\rvecb{k}_{M_2}},\vecb{E}_{-\rvecb{k}_{M_1}},\vecb{E}_{-\rvecb{k}_{M_2}})$.  As determined from its character under $C_{4v}$ (Table \ref{table:PG_char_table_2}),  $S^{M}=E\oplus A^{\prime\prime}_1\oplus B^{\prime\prime}_2$.  The doubly degenerate IRREP $E$ represents the second and third frequency bands as they are degenerate at the $M$-point as shown in Figure \ref{fig:sq_TM_band_lattice}(a), while the $B^{\prime\prime}_2$ and $A^{\prime\prime}_1$ IRREPs represent the first and fourth frequency bands, respectively, as a result of the lattice being centered on an air hole.  Approximate forms for the modes of these bands at the $M$-point are calculated by projecting the symmetry basis onto the three IRREP spaces.  With the origin centered at point $d$, the band-edge modes are:

\begin{equation}
\label{eq:sq_M_VB_modes_d}
M1_{d} = \hat{z} 
\begin{pmatrix}
\cos(\rvecb{k}_{M_1}\cdot\vecb{r}^{d}_{\perp})-\cos(\rvecb{k}_{M_2}\cdot\vecb{r}^{d}_{\perp}) 
\end{pmatrix},
\end{equation}

\begin{equation}
\label{eq:sq_M_E_modes_d}
M(2,3)_{d} = \hat{z} 
\begin{pmatrix}
\sin(\rvecb{k}_{M_1}\cdot\vecb{r}^{d}_{\perp})+\sin(\rvecb{k}_{M_2}\cdot\vecb{r}^{d}_{\perp}) \\
\sin(\rvecb{k}_{M_1}\cdot\vecb{r}^{d}_{\perp})-\sin(\rvecb{k}_{M_2}\cdot\vecb{r}^{d}_{\perp}) 
\end{pmatrix}.
\end{equation}

\begin{equation}
\label{eq:sq_M_HB_modes_d}
M4_{d} = \hat{z} 
\begin{pmatrix}
\cos(\rvecb{k}_{M_1}\cdot\vecb{r}^{d}_{\perp})+\cos(\rvecb{k}_{M_2}\cdot\vecb{r}^{d}_{\perp}) 
\end{pmatrix}.
\end{equation}

\noindent As was done for the $X$ point, shifted band-edge modes for points $e$ and $f$ are easily found and are labeled $M1_{e}$, $M(2,3)_{e}$, $M4_{e}$, and $M1_{f}$, $M(2,3)_{f}$, and $M4_{f}$.    

\subsubsection{Donor modes}
\label{subsubsec:sq_c_donor}
For the square PC lattice, as described earlier, two band-edge minima from which donor modes can be formed are the $X$-point in the second frequency band and the $M$-point in the fourth frequency band (Figure \ref{fig:sq_TM_band_lattice}(a)).  For a symmetric defect centered at points $d$ and $f$ the band-edge modes at the $X$ and $M$ points couple to form resonant modes which transform as IRREPs of $C_{4v}$, whereas for the lower symmetry point $e$ the defect modes transform as IRREPs of $C_{2v,\sigma_v}$.  The representations describing how the $X2_{d}$, $X2_{e}$, and $X3_{f}$ symmetry bases transform under the appropriate point group are given by $S^{d,d_{X2}}$, $S^{e,d_{X2}}$, and $S^{f,d_{X2}}$, respectively.  Similarly, the representations describing how the $M4_{d}$, $M4_{e}$, and $M4_{f}$ symmetry bases transform under the appropriate point group are given by $S^{d,d_{M4}}$, $S^{e,d_{M4}}$, and $S^{f,d_{M4}}$, respectively.   From their characters in Table \ref{table:PG_char_table_2} we find that $S^{d,d_{X2}}=A^{\prime\prime}_1\oplus B^{\prime\prime}_1$, $S^{e,d_{X2}}=A_1\oplus B_2$, and $S^{f,d_{X2}}=E$, while $S^{d,d_{M4}}=A^{\prime\prime}_1$, $S^{e,d_{M4}}=B_2$, and $S^{f,d_{M4}}=B^{\prime\prime}_2$\footnote{The symmetries and fundamental momentum components of the possible donor modes formed from the $M4$ band-edge are trivially given by a single band-edge mode.  This is due to the fact that the $M$-point in the square lattice is highly symmetric, and the group of the wave vector mixes all four of the $M$-points of the IBZ.}.  For points $d$ and $f$, defects may also be formed with lower symmetry than the $C_{4v}$ symmetry of the square lattice.  Using the compatibility relations between the IRREPs of the full and reduced symmetry groups, we determine the new mode structure for a defect of $C_{2v,\sigma_v}$ symmetry (i.e., with mirror planes along the $\hat{x}$ and $\hat{y}$ directions of Figure \ref{fig:sq_TM_band_lattice}(b)).  Projecting the different symmetry bases onto the IRREPS of the point group symmetry of the different cavities one obtains approximate forms for the TM donor modes in the square lattice, the results of which are tabulated in Table \ref{table:sq_lattice_TM_donor_modes}.

\renewcommand{\arraystretch}{1.2}
\renewcommand{\extrarowheight}{0pt}
\begin{table}
\begin{center}
\caption{Symmetry and dominant Fourier components for the $\vecb{E}$-field of TM donor modes formed from the $X2$ and $M4$ band-edges in a square lattice.}
\label{table:sq_lattice_TM_donor_modes}
\begin{tabular*}{\linewidth}{cccc}
\hline
\hline
Cavity Center/Symmetry & Band-Edge & Mode \\
\hline 
$d/C_{4v}$ & $X2$ & $\vecb{E}_{A^{\prime\prime}_1}^{d,d_{X2}} = \hat{z} \bigl(\cos(\rvecb{k}_{X_1}\cdot\vecb{r}^{d}_{\perp})+\cos(\rvecb{k}_{X_2}\cdot\vecb{r}^{d}_{\perp})\bigr)$ \\
$d/C_{4v}$ & $X2$ & $\vecb{E}_{B^{\prime\prime}_1}^{d,d_{X2}} = \hat{z} \bigl(\cos(\rvecb{k}_{X_1}\cdot\vecb{r}^{d}_{\perp})-\cos(\rvecb{k}_{X_2}\cdot\vecb{r}^{d}_{\perp})\bigr)$ \\ 
$e/C_{4v}$ & $X2$ & $\vecb{E}_{A1}^{e,d_{X2}} = \hat{z} \bigl(\cos(\rvecb{k}_{X_2}\cdot\vecb{r}^{e}_{\perp})\bigr)$ \\
$e/C_{4v}$ & $X2$ & $\vecb{E}_{B2}^{e,d_{X2}} = \hat{z} \bigl(\sin(\rvecb{k}_{X_1}\cdot\vecb{r}^{e}_{\perp})\bigr)$ \\
$f/C_{4v}$ & $X2$ & $\vecb{E}_{E,1}^{f,d_{X2}} = \hat{z} \bigl(\sin(\rvecb{k}_{X_2}\cdot\vecb{r}^{f}_{\perp})\bigr)$ \\
$f/C_{4v}$ & $X2$ & $\vecb{E}_{E,2}^{f,d_{X2}} = \hat{z} \bigl(\sin(\rvecb{k}_{X_1}\cdot\vecb{r}^{f}_{\perp})\bigr)$ \\
\hline
$d/C_{2v,\sigma_v}$ & $X2$ & $\vecb{E}_{A^{\prime\prime}_1}^{d,d_{X2}}  \rightarrow  \vecb{E}_{A_1}^{d,d_{X2},1}$ \\
$d/C_{2v,\sigma_v}$ & $X2$ & $\vecb{E}_{B^{\prime\prime}_1}^{d,d_{X2}}  \rightarrow  \vecb{E}_{A_1}^{d,d_{X2},2}$ \\ 
$f/C_{2v,\sigma_v}$ & $X2$ & $\vecb{E}_{E,1}^{f,d_{X2}}  \rightarrow  \vecb{E}_{B_1}^{f,d_{X2}}$ \\
$f/C_{2v,\sigma_v}$ & $X2$ & $\vecb{E}_{E,2}^{f,d_{X2}}  \rightarrow  \vecb{E}_{B_2}^{f,d_{X2}}$ \\
\hline
$d/C_{4v}$ & $M4$ & $\vecb{E}_{A_1^{\prime\prime}}^{d,d_{M4}} = \hat{z} \bigl(\cos(\rvecb{k}_{M_1}\cdot\vecb{r}^{d}_{\perp})+\cos(\rvecb{k}_{M_2}\cdot\vecb{r}^{d}_{\perp}) \bigr)$ \\
$e/C_{2v,\sigma_v}$ & $M4$ & $\vecb{E}_{B_2}^{e,d_{M4}} = \hat{z} \bigl(\sin(\rvecb{k}_{M_1}\cdot\vecb{r}^{e}_{\perp})-\sin(\rvecb{k}_{M_2}\cdot\vecb{r}^{e}_{\perp}) \bigr)$ \\
$f/C_{4v}$ & $M4$ & $\vecb{E}_{B_2^{\prime\prime}}^{f,d_{M4}} = \hat{z} \bigl( \cos(\rvecb{k}_{M_1}\cdot\vecb{r}^{f}_{\perp})-\cos(\rvecb{k}_{M_2}\cdot\vecb{r}^{f}_{\perp}) \bigr)$ \\
\hline
$d/C_{2v,\sigma_v}$ & $M4$ & $\vecb{E}_{A_1^{\prime\prime}}^{d,d_{M4}}  \rightarrow  \vecb{E}_{A_1}^{d,d_{M4}}$ \\
$f/C_{2v,\sigma_v}$ & $M4$ & $\vecb{E}_{B_2^{\prime\prime}}^{f,d_{M4}}  \rightarrow  \vecb{E}_{A_2}^{f,d_{M4}}$ \\
\hline
\hline
\end{tabular*}
\end{center}
\end{table}
\renewcommand{\arraystretch}{1.0}    

\section{FDTD Simulations}
\label{sec:FDTD_init}

In this section, FDTD simulations of the donor defect modes identified in Section \ref{sec:GT} are presented.  The FDTD calculations establish the effectiveness of the chosen symmetry bases of the group theoretical analysis in describing the defect cavity modes.  To reduce computational time, we limit our study to 2D FDTD simulations with an effective index used to account for vertical guiding; previous papers\cite{ref:Painter12,ref:Painter13} have shown that 3D FDTD simulations of TE-like modes in full PC slab WG defect cavities correspond well with the modes found through the approximate group theory analysis.  
The FDTD calculations were performed on a mesh with 20 points per lattice spacing.  Excitation of the cavity modes was performed by an initial field ($E_z$) with a localized Gaussian profile, located in a position of low symmetry so as not to exclude any possible resonant modes. TM modes are selected out by constraining the normal component of the field to be purely $E_{z}$, and a pair of mirror planes ($\sigma_x$, $\sigma_y$) were used to filter out cavity modes according to their projection on the IRREPs of $C_{2v}$.  Each FDTD-generated cavity mode is thus labelled by the $C_{2v}$ IRREP by which it transforms and an index corresponding to its energy (frequency) level.

FDTD-generated results are compared with the results of the symmetry analysis both through confirmation of the symmetry classifications under the IRREPs of $C_{2v}$ (although a careful analysis of all available mirror planes within the FDTD simulations would allow classification under higher symmetries such as $C_{4v}$ or $C_{6v}$) and through an analysis of the dominant Fourier components of the modes (through examination of the spatial Fourier transform of $E_{z}$ ($\widetilde{E}_{z}
$)).     
\subsection{Hexagonal Lattice}
\label{subsec:hex_lattice_FDTD}

\begin{figure}[ht]
\begin{center}
\epsfig{figure=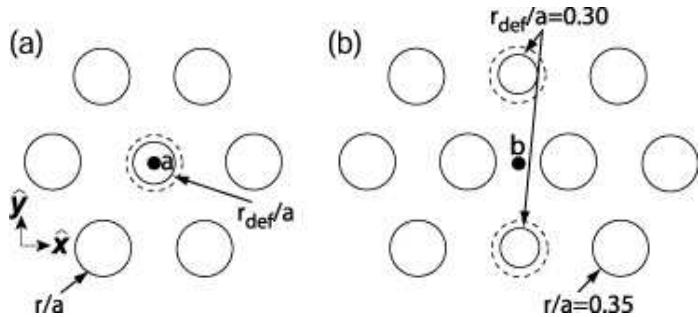, width=0.7\linewidth}
\caption{Defect geometries for donor modes about the (a) $a$-point and (b) $b$-point.}
\label{fig:hex_defects} 
\end{center}
\end{figure}       

\begin{figure}[ht]
\begin{center}
\epsfig{figure=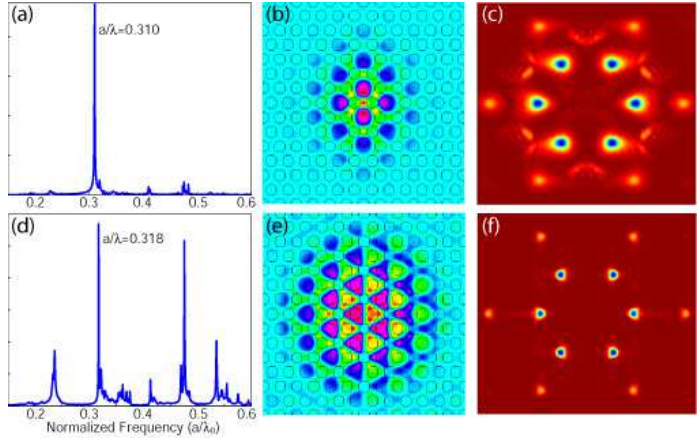, width=0.7\linewidth}
\caption{Donor modes about the $b$-pt. in the hexagonal lattice. (a) Spectrum for $A_1$ symmetry modes. (b) $E_z$ and  (c) $|\widetilde{E}_z|$ for $A_1$ mode.  (d) Spectrum for $B_1$ symmetry modes. (c) $E_z$ and (d) $|\widetilde{E}_z|$ for $B_1$ mode.}
\label{fig:hex_b_pt} 
\end{center}
\end{figure}

We begin by considering donor-type defect modes in the hexagonal lattice formed around the $J$-point of the third frequency band, where from our symmetry analysis we expect to see modes of $A_1$ and $B_1$ symmetries (under $C_{2v}$).  In Figure \ref{fig:hex_defects} we show two examples of simple defects within the hexagonal lattice, one localized around the point $a$ of the lattice, the other around point $b$.  The chosen defect around the $b$-point in the hexagonal lattice, the defect cavity we will consider here, consists of two reduced size air holes along the $y$-axis, with $r/a=0.35$ and $r_{\text{def}}/a=0.30$ (Figure \ref{fig:hex_defects}(b)).  The spectrum for $A_1$ modes shown in Figure \ref{fig:hex_b_pt}(a) shows a single dominant peak in the frequency range of interest; the real space and Fourier space versions of this field are given in Figure \ref{fig:hex_b_pt}(b)-(c), and indicate that the mode has dominant Fourier components at the six $J$-points within the IBZ, as predicted by our symmetry analysis.  The Fourier transformed electric field for the $B_1$ symmetry mode is shown in Figure \ref{fig:hex_b_pt}(f); the mode has its dominant Fourier components at the six $J$-points, and is somewhat delocalized in comparison to the $A_1$ mode (Figure \ref{fig:hex_b_pt}(e)).  Similar results and correspondence were found for the defect around point $a$ of the lattice (Figure \ref{fig:hex_defects}(a)).  As a final comment, we note that FDTD-generated spectra for $A_2$ and $B_2$ symmetry modes were also examined (for both points $a$ and $b$), and significant resonance peaks were not seen in the frequency range of interest.   

\subsection{Square Lattice}
\label{subsec:sq_lattice_FDTD}

In Section \ref{sec:GT} we identified three high symmetry points in the square lattice about which defect modes are formed.  For each of these points, we considered donor modes off the $X$ point of the second frequency band and $M$ point of the fourth frequency band.  We now present FDTD simulations which verify the symmetry classification and Fourier space composition for the defect modes predicted by the group theory analysis.

\begin{figure}[ht]
\begin{center}
\epsfig{figure=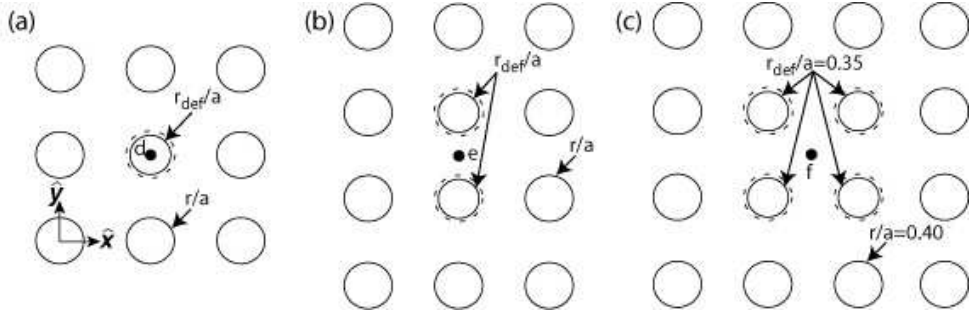, width=\linewidth}
\caption{Defect geometries for donor modes about the (a) $d$-point and (b) $e$-point and (c) $f$-point.}
\label{fig:sq_defects_donor} 
\end{center}
\end{figure}         

\begin{figure}[ht]
\begin{center}
\epsfig{figure=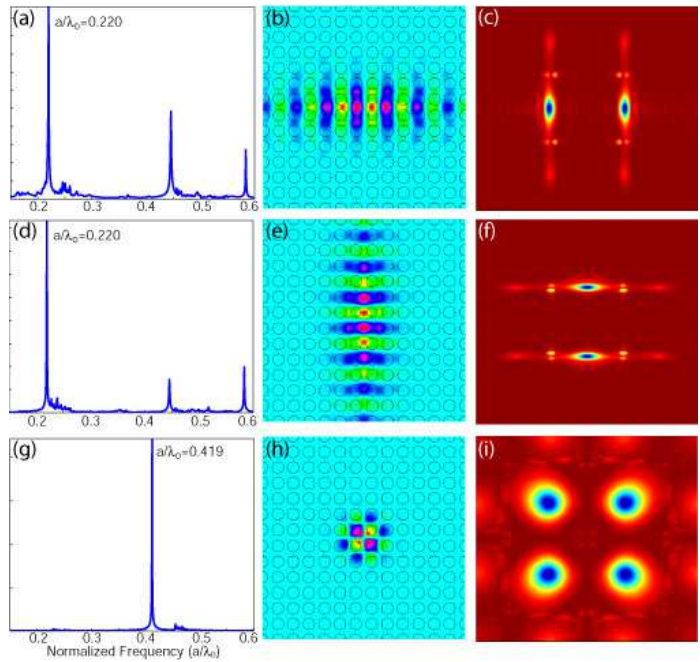, width=0.7\linewidth}
\caption{Properties of $f$-pt. modes in the square lattice.  (a) $B_1$ symmetry mode spectrum. (b) $E_z$ and (c) $|\widetilde{E}_z|$ for low frequency $B_1$ mode. (d)  $B_2$ symmetry mode spectrum. (e) $E_z$ and (f) $|\widetilde{E}_z|$ for low frequency $B_2$ mode. (g) $A_2$ symmetry mode spectrum. (h) $E_z$  and (i) $|\widetilde{E}_z|$ for high frequency $A_2$ mode.}
\label{fig:sq_f_pt_modes} 
\end{center}
\end{figure}       

Figure \ref{fig:sq_defects_donor} shows three examples of defect cavities formed about the $d$, $e$, and $f$ points of the square lattice.  We present here the FDTD results for the defect cavity about the $f$ point.  In this cavity a localized defect state is formed that has $\hat{z}$-component of the electric field predominantly within the high refractive index material, which for application as a QC laser cavity would provide the largest possible overlap factor with the gain region in the semiconductor.  Donor modes about the $f$-point are created by reducing the size of four air holes (Figure \ref{fig:sq_defects_donor}(c)).  For this example, we have taken $r/a=0.40$ and $r_{\text{def}}/a=0.35$.  The symmetry classifications for the modes predicted in Section \ref{sec:GT} are confirmed as this perturbation to the dielectric lattice creates donor modes of type $B_1$, $B_2$, and $A_2$ symmetry, with the primary $B_1$ mode at a normalized frequency of $a/\lambda_0=0.220$ (Figure \ref{fig:sq_f_pt_modes}(a)), the dominant $B_2$ mode at a normalized frequency of $a/\lambda_0=0.220$ (Figure \ref{fig:sq_f_pt_modes}(d)), and the main $A_2$ mode at $a/\lambda_0=0.419$ (Figure \ref{fig:sq_f_pt_modes}(g)).  The real and Fourier space versions of the electric field for the low frequency peaks show that the $B_1$ mode (Figure \ref{fig:sq_f_pt_modes}(b)-(c)) has dominant Fourier components at ${\pm}X_2$, and the $B_2$ mode (Figure \ref{fig:sq_f_pt_modes}(e)-(f)) has its dominant Fourier components at ${\pm}X_1$, consistent with the group theory analysis.  These modes are in fact degenerate (consistent with the group theory prediction that they originate from a 2D IRREP for a defect of $C_{4v}$ symmetry) and act as a dipole-like pair of modes, in analogy with the hexagonal lattice dipole modes studied in previous work\cite{ref:Painter12}.  Finally, $|\rvecb{E}|$ and $|\widetilde{\rvecb{E}}_z|$ for the higher frequency $A_2$ peak are shown in Figure \ref{fig:sq_f_pt_modes}(h)-(i), and show that it is composed of dominant Fourier components at the four $M$ points.      

\section{Summary and Discussion}
\label{sec:summary}

The group theory analysis presented in Section \ref{sec:GT} can be used to ascertain a number of important properties about TM-polarized defect modes in connected PC lattices, and as such, serves as a starting point from which more careful designs of PC defect cavities can originate.  For example, recognizing potential modal degeneracies and modifications that can be made to split or remove these degeneracies can be important in laser design\cite{ref:Park1,ref:Painter12}.  Coupling of the cavities to other in-plane elements can benefit from an understanding of the modes' dominant in-plane emission directions, while knowledge of the position of field nodes/antinodes with respect to the dielectric and air regions is important as some applications (such as coupling to semiconductor quantum dots) benefit from field maxima in the dielectric regions, while others (such as coupling to states of cold atoms ``dropped'' within the cavity) are better served by field maxima in the air holes.  

With the results obtained from the group theory analysis in hand, more sophisticated modeling, such as full 3D FDTD simulations, can be used to determine more quantitative properties of cavity modes such as modal frequency, cavity quality factor ($Q$), and precise near and far-field emission pattern.  This approach was used in Ref. \cite{ref:Srinivasan1}, which focused on the design of high-$Q$ wavelength-scale PC microcavities.  In this work, the group theory based approach was used to analyze TE-polarized modes localized to different high symmetry points within connected square and hexagonal lattice PCs.  Fourier space design principles were then used to determine which of these candidate modes were particularly suitable for high-$Q$ applications, and then full 3D FDTD modeling was used in the final design steps.  

\begin{figure}[ht]
\begin{center}
\epsfig{figure=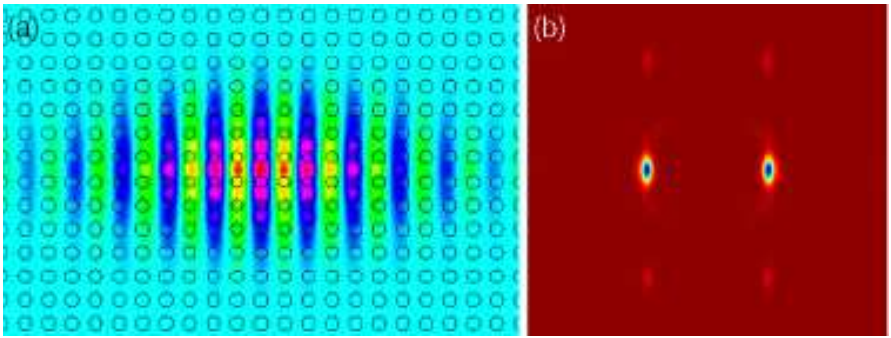, width=0.9\linewidth}
\caption{Properties of $X$-point $\vecb{E}_{A_1}^{e,d_{X2}}$ symmetry mode in a graded square lattice.  (a) $E_z$ (b) $|\widetilde{E}_z|$.}
\label{fig:sq_lattice_graded} 
\end{center}
\end{figure}       

As a simple example, we conclude by considering the formation of a PC defect laser cavity for TM-polarized modes (used in conjunction with QC heterostructures, for example).  The square lattice defect cavities considered in Section \ref{sec:GT} had fairly large-size air holes ($r/a=0.45$), which in fabricated devices can be problematic for a number of reasons, including reduced dielectric-field overlap ($\Gamma$) and poor thermal stability due to the porosity of the lattice.  In electrically injected devices, this porosity could also hinder defect mode lasing, as pumping of the central (defect) region of the cavity will be more difficult for such small current paths.  On the other hand, larger holes create larger frequency-space gaps in which the modes can be localized.  One method for achieving high in-plane reflectivity while reducing the porosity of the lattice is to employ a graded lattice design\cite{ref:Srinivasan1}.  Starting with the requirement that our lasing mode have an electric field antinode in a dielectric region (this mode is similar to the $\vecb{E}_{E,1}^{f,d_{X2}}$ mode of the $f$-point defect cavity), we use our group theory based analysis to select the low-frequency $\vecb{E}_{A_1}^{e,d_{X2}}$ mode as a mode of interest.  We then employ the graded lattice geometry shown in Figure \ref{fig:sq_lattice_graded}(a) to create such a mode.  It consists of two levels of confinement: first, an initial jump in hole radius between the holes centered at $(0,\pm{a/2})$ (which have $r/a=0.225$) and their nearest neighbors (which have $r/a=0.25$), and then, a parabolic increase in hole radius moving outwards from the defect center (with a maximum value $r/a=0.30$, equal to the hole size used in lasing devices in Ref. \cite{ref:Colombelli2}).  The initial jump in $r/a$ serves as a potential well to help confine the mode in real space, while the grade allows for the hole radius to be kept small in the region where the mode is primarily located (to keep $\Gamma$ high), but large enough outside of this region to produce a high in-plane reflectivity.  The resulting real space field $E_z$ and Fourier transform $|\widetilde{E}_z|$ for the $A_1$ mode are shown in Figure \ref{fig:sq_lattice_graded}.  Based on these images, we see that our initial objectives have been met: the mode remains well localized to the central region of the cavity, despite having now chosen considerably smaller holes.  With this lattice geometry taken as a template, the Fourier space design principles of Ref. \cite{ref:Srinivasan1} could now be applied, in conjunction with 3D FDTD simulations, to optimize the cavity $Q$ factor for use in high performance devices.         

\ack{K. Srinivasan thanks the Hertz Foundation for its graduate fellowship support.}  
\\

\bibliography{./PBG}

\end{document}